\documentclass[10pt,journal,doublecolumn,twoside]{IEEEtran}
\IEEEoverridecommandlockouts

\usepackage{subfigure}
\usepackage{colortbl}
\usepackage{bm}

\usepackage{cite}
\usepackage{amsmath,amssymb,amsfonts}
\usepackage{algorithmic}
\usepackage{algorithm}
\usepackage{graphicx}
\usepackage{textcomp}
\usepackage{mathrsfs}
\usepackage{bm}
\usepackage{booktabs}
\usepackage{graphicx}
\usepackage{rotating}
\usepackage{subeqnarray}
\usepackage{cases}

\usepackage{graphicx}  
\usepackage{url}       
\usepackage{rotating}
\usepackage{amsmath}   
\usepackage{cite}

\usepackage{amsfonts,amssymb}

\usepackage{stfloats}

\usepackage{cases}
\usepackage{algorithm}
\usepackage{multirow}
\usepackage{algorithmic}

\newcommand{\myvec}[1]%
{\stackrel{\raisebox{-2pt}[0pt][0pt]
{\small$\rightharpoonup$}}{#1}}

\newcommand{\ls}[1]
    {\dimen0=\fontdimen6\the\font
     \lineskip=#1\dimen0
     \advance\lineskip.5\fontdimen5\the\font
     \advance\lineskip-\dimen0
     \lineskiplimit=.9\lineskip
     \baselineskip=\lineskip
     \advance\baselineskip\dimen0
     \normallineskip\lineskip
     \normallineskiplimit\lineskiplimit
     \normalbaselineskip\baselineskip
     \ignorespaces
    }

\usepackage[left=0.68in,right=0.70in,top=0.72in,bottom=0.58in]{geometry}
\begin{document}

\bibliographystyle{IEEEtran}
\setlength{\columnsep}{0.24in}
\title{Fog Based Computation Offloading for\\ Swarm of Drones
\thanks{This work was supported in part by National Key R\&D Program of China(No. 2016YFE0123000, 2017YFE0121400 and 2016YFE0207000), the National Natural Science Foundation of China (No. 61571338, U1709218, 61771368 and 61672131), the Young Elite Scientists Sponsorship Program by CAST(No. 2016QNRC001), and the Young Talent Support Fund of Science and Technology of Shaanxi Province(No. 2018KJXX-025)}
}
\author{\IEEEauthorblockN{Xiangwang Hou,
Zhiyuan Ren, Wenchi Cheng, Chen Chen and
Hailin Zhang
}\\
\IEEEauthorblockA{State Key Laboratory of Integrated Services Networks,
Xidian University, Xi'an, China}\\
E-mail: \{\emph{xwhou@s-an.org}, \{\emph{zyren, wccheng, cc2000 and hlzhang\emph}\}\emph{@xidian.edu.cn\emph}\}
}

\maketitle
\thispagestyle{empty}
\pagestyle{empty}
\begin{abstract}
Due to the limited computing resources of swarm of drones, it is difficult to handle computation-intensive tasks locally, hence the cloud based computation offloading is widely adopted. However, for the business which requires low latency and high reliability, the cloud-based solution is not suitable, because of the slow response time caused by long distance data transmission. Therefore, to solve the problem mentioned above, in this paper, we introduce fog computing into swarm of drones (FCSD). Focusing on the latency and reliability sensitive business scenarios, the latency and reliability is constructed as the constraints of the optimization problem. And in order to enhance the practicality of the FCSD system, we formulate the energy consumption of FCSD as the optimization target function, to decrease the energy consumption as far as possible, under the premise of satisfying the latency and reliability requirements of the task. Furthermore, a heuristic algorithm based on genetic algorithm is designed to perform optimal task allocation in FCSD system. The simulation results validate that the proposed fog based computation offloading with the heuristic algorithm can complete the computing task effectively with the minimal energy consumption under the requirements of latency and reliability.
\end{abstract}
\vspace{3mm}

\section{Introduction}
Swarm of drones, which are considered as an intensely promising development direction of Unmanned Aerial Vehicles (UAVs), has made great progress in recent years. Swarm of drones are composed by numerous small and low-cost UAVs, through collaborating with each other, the drones can show strong ability to accomplish the tasks which are difficult for a single large UAV. As a consequence, swarm of drones are widely used for a variety of applications, such as agriculture, smart city, search and rescue, remote sensing, military, etc \cite{wurenjiyingyong2}. For most of these applications, drones are brought to deal with computation-intensive tasks, such as path planning, pattern recognition, etc\cite{renwujvli}. However, due to its limited resources (e.g., battery power, computing capability), the single drone is too difficult to handle the complicated task locally\cite{messous2017computation}. Therefore, to address the computation-intensive tasks mentioned above, some researches considered computation offloading to a cloud server, and then obtain the result from the cloud\cite{jisuanxiezaicloud}. In this manner, the capability of swarm of drones is greatly enhanced in a virtual way. And it is suitable for some business (e.g., topographic mapping, resource exploration, environmental monitoring, etc.) which are not sensitive to latency and reliability. But in practice, quite a few computing tasks which the drones need to process, have low latency and high reliability requirements, such as military object recognition, disaster rescue, emergency obstacle avoidance, etc \cite{liu2018data}\cite{guo2019deep}. However, cloud servers are generally located far away from the drones, and long-distance data transmission will lead to high latency, even in some harsh environments, there is no working wireless infrastructure to connect the drones and cloud. Hence, the cloud based computation offloading is not suitable to address the latency and reliability sensitive business.

In order to further enhance the ability of swarm of drones to cope with computation-intensive tasks, specifically focusing on those tasks with low latency and high reliability requirements, we introduce fog computing \cite{fogheUAV} into swarm of drones. The drones which close to the initiator drone are thought to be fog computing nodes to complete the computing task collaboratively.

Fog computing is a novel computing paradigm which is not intended to replace cloud computing but to compliment it. Recently, there are many researches about fog computing enhancing cloud computing. In \cite{chuanshusulv}, authors considered leveraging buses as fog computing servers to provide fog computing services for the mobile users on bus and share the pressure of roadside cloudlets. The authors in \cite{souza2016handling} proposed combined fog-cloud architecture to reduce the latency of service. In \cite{shah2018hierarchical}, authors formulated a computation offloading game to improve the quality of experience of IoT users in hierarchical fog-cloud computing architecture. But there is no existing research to introduce fog computing into the work of swarm of drones, as a supplementary solution for the computation offloading of cloud computing.

In practice, swarm of drones usually work in harsh environments, and inevitable disturbances (e.g., hardware damage, software breakdown, communication link failure, etc.) will lead to the failure of the task. Hence, besides considering the latency guarantee, a proper reliability-guarantee mechanism is especially needed. However, there are few existing researches about fog (or edge) based computation offloading both considering the latency and reliability guarantee\cite{kekaoxingmeirenzuo}. Therefore, in order to complete the computing task within low latency and high reliability requirements, we construct a certain mathematical model for system's latency and reliability during task execution, and take the latency and reliability as the constraints of the optimization problems formulated, thus the computation offloading scheme must be able to meet the requirements of the business on latency and reliability.

A big challenge for swarm of drones utilizing fog computing to deal with computing task is their limitation of battery endurance, hence, under the premise of ensuring the completion of the task within low latency and high reliability requirements, we formulated the energy consumption of whole swarm of drones as the target function of optimization problem, to reduce the overall energy consumption and extend the working time of the swarm of drones as far as possible. Since the formulated problem is NP-hard, we design a heuristic algorithm based on genetic algorithm to solve the problem.

In summary, the main contributions of this paper are as follows:
\begin{itemize}
  \item To enhance the ability of swarm of drones to handle complicated tasks, we introduce fog computing into swarm of drones (FCSD), as a supplementary solution for the cloud based computation offloading.
  \item To meet the latency and reliability requirements of the computing tasks, we construct a certain mathematical model for system latency and reliability during task execution.
  \item To improve the practicality of the FCSD system in practice, the energy consumption of FCSD is constructed as the optimization target function, to reduce the energy consumption so that extend the working time of swarm of drones.
  \item To solve the NP-hard problem formulated, we design a latency and reliability constrained minimum energy consumption algorithm based on genetic algorithm (LRGA-MIE).
\end{itemize}


The rest of the paper is organized as follows. The system model and problem formulation are presented in Section II. Section III presents the offloading algorithm proposed. The simulation results and analyses are given in Section IV. Finally, Section V conclude the paper.
%

\section{SYSTEM MODEL AND PROBLEM FORMULATION}
\begin{figure}[htb]
  \centering
  \includegraphics[width=7.5cm]{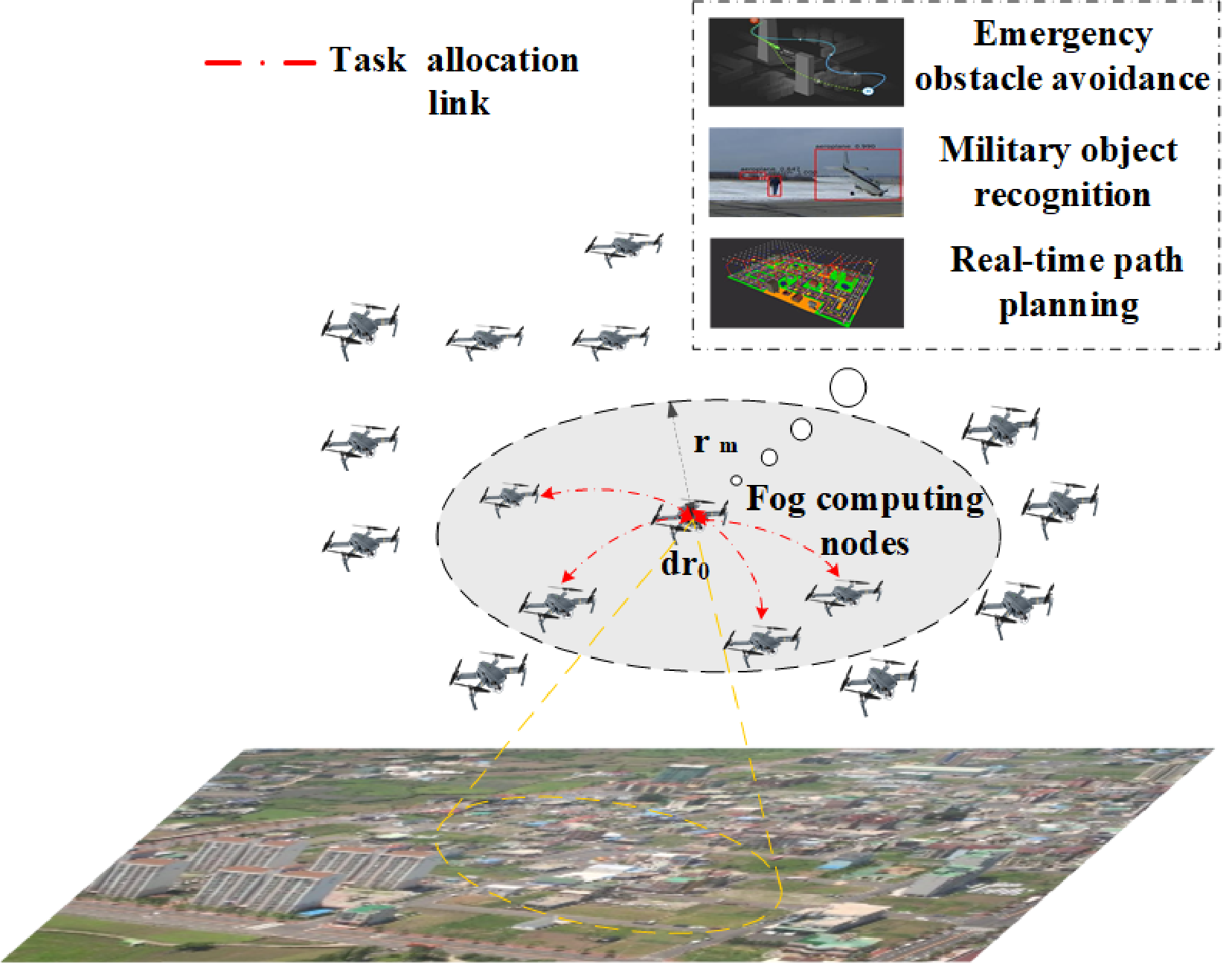}
 \caption{The computation offloading architecture of FCSD}\label{drones}
 \vspace{-3mm}
\end{figure}

In order to enhance the ability of swarm of drones addressing computation-intensive tasks which are sensitive to latency and reliability, the FCSD system is proposed. The architecture of FCSD is shown in Fig. \ref{drones}.

The drone $dr_0$ has a computing task ${\varPsi_0}\triangleq\{D_0,\alpha_0,T_0,R_0\}$, where $D_0$ denotes the input size of the total task; $T_0$ and $R_0$ represent the latency and the reliability constraints, respectively. We define $E_0$ as the total required amount CPU cycles to complete the task $\varPsi_0$. The number of CPU cycles $E_0$ is modeled as $ E_0=\alpha_0D_0\ $, where $\alpha_0$($\alpha_0 > 0$) depends on the computational complexity of the task\cite{jisuanfuzadu2}. The drone $dr_0$ requests nearby drones $dr_i$ that can serve as the fog nodes to complete the task $\varPsi_0$ collaboratively. These drones available nearby, denoted by a set $\mathcal{D}=\left\{dr_1,dr_2,\ldots,dr_p \right\}$, are equipped with storage and computation resources. We define $f_0$ as the CPU frequency of the drone $dr_0$. Similarly, the CPU frequency of the drones available nearby, denoted by a set $\mathcal{F}=\left\{f_1,f_2,\ldots,f_p\right\}$. The coordinate of the drone $dr_0$ is $\left(x_0,y_0,z_0\right)$. The
$\mathcal{C}=\left\{(x_1,y_1,z_1),(x_2,y_2,z_2),\cdots,(x_p,y_p,z_p)\right\}$  are the three-dimensional coordinates of the drones available nearby, respectively.

The distance between the drone $dr_i \in \mathcal{D}$ and the drone $dr_0$ can be given by
\begin{eqnarray}
\resizebox{0.85\hsize}{!}{$
\label{eq1}
  g_{0,i}={{{[{({x_0}-{x_i})^{2}}+{({y_0}-{y_i})}^{2}+{({{z_0}-{z_i}})}^{2}]}}^{\frac{1}{2}}},  g_{0,i}\le r $},
\end{eqnarray}
where $r$ is the maximum communication radius of individual drones.
According to \cite{chuanshusulv}\cite{cheng2016statistical}, the uplink rate from $dr_0$ to $dr_i$ can be given as
\begin{eqnarray}
\resizebox{0.7\hsize}{!}{$
\label{eq2}
  R^{\textrm{UL}}{(0,i)} = W^{\textrm{UL}}\log_2\left( 1+\frac{{P_{\textrm{Tx}}}\left({g_{0,i}}^{-\gamma}{\left|h_0     \right|}\right)}{N_0}\right)$},
\end{eqnarray}
where $W^{\textrm{UL}}$ represents the uplink channel bandwidths between the drone $dr_0$ and $dr_i$; ${P_{\textrm{Tx}}}$ denotes the transmission power of the drone $dr_0$; $\gamma$ is the path loss exponent which ranges from $2 \le {\gamma} \le 5$ ; $h_0$ is the complex Gaussian channel coefficient which follows the complex normal distribution CN(0,1); $N_0$ is the additive white Gaussian noise(AWGN).

The task $\varPsi_0$ would be partition into several subtasks by the drone $dr_0$ and distributed to multiple drones. In practice, how to partition a task depends on not only application, but also the requirements, which is worth studying further. Therefore, for simplicity, it can be assumed that the task can be divided into any proportion with arbitrary precision, and there is no overlap between any two subtasks\cite{kekaoxingmeirenzuo}. According to the status of the $dr_0$ and $dr_i \in \mathcal{D}$, the initiator drone $dr_0$ can make different task offloading and allocation decisions\footnote{In practice, the drone interacts with the surrounding drones to complete the formation, networking, collaborative works, etc., which enables the drone to be aware of the status of the surrounding drones.\cite{khan2017flying}}. We define $\rho ~(0 \leq \rho \leq 1) $ as the offloading coefficient, therefore, the part of the task $\varPsi_0$ which need to be executed locally, can be described as $\rho{\varPsi_0}$, and the part of the task which need to be offloaded to the drones available nearby is $\left(1-\rho\right){\varPsi_0}$. Then, we denote the subtask offloaded to the drone $dr_i$ as $\lambda_i(1-\rho){\varPsi_0}$, where $\lambda_i \in [0,1]$, and $\sum\limits_{i=1}^p{\lambda_i}=1$. We define
 $\bm{\lambda} = [\lambda_1,\lambda_2,\ldots,\lambda_p]^\textrm{T}$ as the task allocation vector.


After decision, the drone $dr_0$ and the drones $dr_i \in \mathcal{D}$ are orchestrated to perform distributed computing to complete the task $\varPsi_0$ collaboratively. These low-cost drones fly slowly and tend to form a relatively stable formation, rather than constantly changing\cite{wurenjiyingyong2}. In the meantime, the transmission latency of the subtask is extremely short when the data size is small. Therefore, we assume that the relative distance and the states of the drones will not change during the task assignment process, and each assigned subtask will be executed immediately on the drone $dr_i$ and $dr_0$.

\vspace{-0.3cm}
\subsection {Latency Model}
The latency of the drone $dr_0$ processing the subtask $\rho\varPsi_0$ is defined as
\vspace{-0.1cm}
\begin{eqnarray}
\resizebox{0.28\hsize}{!}{$
\begin{split}
\label{eq4}
T_{\textrm{Local}}&=\frac{\rho{\alpha_0}{D_0}}{f_0}.
\end{split}$}
\end{eqnarray}

When the drone $dr_0$ offloading the subtask ${\lambda_i}{(1-\rho)}\varPsi_0$ to the drone $dr_i$, the size of the transmitted data will be ${\beta}{\lambda_i}{(1-\rho)}D_0$, where $\beta ~ (\beta \geq 1)$ represents a ratio of the transmitted data size to the original task data size due to transmission overhead\cite{jisuanfuzadu2}. Thus, the transmission latency of the subtask from the drone $dr_0$ to the drone $dr_i$ is
 \begin{eqnarray}
\resizebox{0.30\hsize}{!}{$
\label{eq5}
{T_i}^{\textrm{UL}}=\frac{{{\beta}{\lambda_i}{(1-\rho)}D_0}}{R^{\textrm{UL}}(0,i)}$}.
 \end{eqnarray}
And the computation latency of the subtask addressed on the drone $dr_i$ is
\vspace{-0.1cm}
\begin{eqnarray}
\resizebox{0.35\hsize}{!}{$
\label{eq6}
 {T_i}^{\textrm{Comp}}=\frac{{{\alpha_0}{\lambda_i}{(1-\rho)}D_0}}{f_i}$}.
\end{eqnarray}
Due to the data size of the result of each subtask is much smaller than the input one, the latency caused by downlink transmission can be neglected\cite{jisuanjieguo1}. The total execution latency of the subtask completed on the drone $dr_i$ is given by
\begin{eqnarray}
\resizebox{0.55\hsize}{!}{$
\begin{split}
\label{eq7}
T_i &= T^{\textrm{UL}}_i + T^{\textrm{Comp}}_i \\
    &= \frac{{{\beta}{\lambda_i}{(1-\rho)}D_0}}{R^{\textrm{UL}}(0,i)} + \frac{{{\alpha_0}{\lambda_i}{(1-\rho)}D_0}}{f_i}.
\end{split}$}
\end{eqnarray}

Therefore, the total execution latency of the task $\varPsi_0$ can be described as
\begin{equation}
\resizebox{0.81\hsize}{!}{$
\begin{split}
\label{eq8}
T_{\textrm{Total}}&={\max\limits_{i\in p}\{ T_{\textrm{Local}},T_i\}}   \\
&={\max\limits_{i\in p}}\bigg\{\frac{\rho{\alpha_0}{D_0}}{f_0},\frac{{{\beta}{\lambda_i}{(1-\rho)}D_0}}{R^{\textrm{UL}}(0,i)} + \frac{{{\alpha_0}{\lambda_i}{(1-\rho)}D_0}}{f_i}\bigg\}.
\end{split}$}
\end{equation}

To meet the latency requirement of the task $\varPsi_0$, the total execution latency $T_{\textrm{Total}}$ should meet the constraint $T_{\textrm{Total}}\le T_0$.
\vspace{-0.5cm}
\subsection{Reliability Model }
The swarm of drones usually work in harsh environments, and inevitable disturbances (e.g., hardware damage, software breakdown, communication link failure, etc.) will lead to the failure of the whole task, and always with serious consequence. Therefore, a proper reliability-guarantee capability is especially needed to ensure the successful completion of the mission.

According to the widely accepted reliability model proposed by Shatz\cite{kekaoxingmoxing}, the system reliability is that `` the product of the probability that each processor is operational during the time of processing the tasks assigned to it, and the probability that each communication link is operational during the period of the data transmission.''The failure of the drones and communication links follow a Poisson process\cite{kekaoxingmoxing}, further, the failure rates of the drone $dr_0$ and $dr_i$ are defined as $\nu_{0}$ and $\nu_{i}$, respectively, and the failure rate of the communication links between $dr_0$ and $dr_i$ is defined as $\mu_{0,i}$. Therefore, the computation reliability of the drone $dr_0$ and $dr_i$ can be represented as ${e}^{{-\nu_i}{\frac{\rho\alpha_0{D_0}}{f_0}}}$ and ${e}^{{-\nu_i}{\frac{\lambda_i(1-\rho)\alpha_0{D_0}}{f_i}}}$, respectively. And the communication reliability between $dr_0$ and $dr_i$ can be represented as ${e}^{{-\mu_{0,i}}{\frac{\lambda_i(1-\rho)\beta{D_0}}{R^{\textrm{UL}}{(0,i)}}}}$. The reliability of the subtask which executed locally can be represented as
\begin{eqnarray}
\resizebox{.3\hsize}{!}{$
\label{eq9}
R_{\textrm{Local}}={e}^{{-\nu_0}{\frac{\rho\alpha_0{D_0}}{f_0}}}$}.
\end{eqnarray}

Then, the reliability of the subtask which distributed to the drone $dr_i$ can be represented as
\begin{eqnarray}
\resizebox{.60\hsize}{!}{$
\label{eq10}
R_{i}={e}^{{{-\nu_i}{\frac{\lambda_i(1-\rho)\alpha_0{D_0}}{f_i}}}{{-\mu_{0,i}}{\frac{\lambda_i(1-\rho)\beta{D_0}}{R^{\textrm{UL}}(0,i)}}}}{.}$}
\end{eqnarray}

Therefore, the reliability of the swarm of drones during the execution time of the task $\varPsi_0$ can be given by
\vspace{-0.2cm}
\begin{eqnarray}
\resizebox{.83\hsize}{!}{$
\begin{split}
\label{eq11}
R_{\textrm{Total}}&=R_{\textrm{Local}}\prod\limits^{p}_{i=1}R_i \\
         &=  {e}^{{{-\nu_0}{\frac{\rho\alpha_0{D_0}}{f_0}}}+\sum\limits_{i=1}^p\left({{{-\nu_i}{\frac{\lambda_i(1-\rho)\alpha_0{D_0}}{f_i}}}{{-\mu_{0,i}}{\frac{\lambda_i(1-\rho)\beta{D_0}}{R^{\textrm{UL}}(0,i)}}}}\right)}.
\end{split}$}
\end{eqnarray}

To meet the reliability requirement of the task $\varPsi_0$, the total reliability $R_{\textrm{Total}}$ should meet the constraint $R_{\textrm{Total}} \geq R_0$.
\vspace{-0.3cm}
\subsection{Energy Consumption Model}
To improve the practicality of the FCSD system, how to minimize the energy consumption, under the premise of ensuring the completion of the task within latency and reliability requirements, must be taken into account.  Therefore, a mathematical model that minimizes the energy consumption of FCSD processing a single task is constructed.
\subsubsection{Computational energy consumption}
The computational energy consumption of the drone $dr_0$ and $dr_i$ can be given by
\vspace{-2mm}
\begin{eqnarray}
\resizebox{.3\hsize}{!}{$
\label{eq12}
E^{\textrm{Comp}}_{\text{Local}} = kf_0^{\sigma}T^{\textrm{Local}}{;}$}
\end{eqnarray}
\vspace{-8mm}
\begin{eqnarray}
\resizebox{.3\hsize}{!}{$
\label{eq12b}
E^{\textrm{Comp}}_{i} = kf_i^{\sigma}T^{\textrm{Comp}}_{i},$}
\end{eqnarray}
%
%
%
respectively, where $kf_0^{\sigma}$ and $kf_i^{\sigma}$ are the computation power of the drone $dr_0$ and $dr_i$. According to \cite{power2}, the $k > 0$ and the $\sigma \geq 2$ (which usually close to 3), are the positive constant. As in \cite{dinh2017offloading}, the $k$ and the $\sigma$ can be set as $1.25 \times 10 ^{-26}$ and $3$, respectively.

Therefore, the total computational energy consumption of the swarm of drones is represented as
\vspace{-0.2cm}
\begin{eqnarray}
\resizebox{.7\hsize}{!}{$
\begin{split}
\label{eq13}
E^{\textrm{Comp}}_{\textrm{Total}}&=kf_0^{\sigma}T_{\textrm{Local}}+ \sum\limits_{i=1}^{p}{kf_i^{\sigma}T^{\textrm{Comp}}_i} \\
&=kf_0^{\sigma}{\frac{\rho{\alpha_0}{D_0}}{f_0}}+ \sum\limits_{i=1}^{p}{kf_i^{\sigma}{ \frac{{{\alpha_0}{\lambda_i}{(1-\rho)}D_0}}{f_i}}}.
\end{split}$}
\end{eqnarray}
\subsubsection{Transmission energy consumption}
The transmission energy consumption of the drone $dr_0$ and the drone $dr_i$ can be given as
\vspace{-2mm}
\begin{eqnarray}
\resizebox{.26\hsize}{!}{$
\label{eq14}
E_{\textrm{Local}}^{\textrm{Trans}} = P_{\textrm{Tx}}T^{\textrm{UL}}_{i};$}
\end{eqnarray}
\vspace{-6.5mm}
\begin{eqnarray}
\resizebox{.26\hsize}{!}{$
\label{eq14b}
E_{\textrm{i}}^{\textrm{Trans}} = P_{\textrm{Rx}}T^{\textrm{UL}}_{i},$}
\end{eqnarray}
respectively, where ${P_{\textrm{Tx}}}$ and ${P_{\textrm{Rx}}}$ denote the transmitting and receiving power of the drone $dr_0$ and $dr_i$, respectively, which are regarded as constant \cite{dinh2017offloading}.
Therefore, the total transmission energy consumption of the FCSD system can be given by
\begin{eqnarray}
\resizebox{.78\hsize}{!}{$
\begin{split}
\label{eq15}
E_{\textrm{Total}}^{\textrm{Trans}}&=\sum\limits^{p}_{i=1}{ E_{\textrm{Local}}^{\textrm{Trans}}}+\sum\limits^{p}_{i=1}{E_{\textrm{i}}^{\textrm{Trans}}} \\
&=\sum\limits^{p}_{i=1}{P_{\textrm{TR}}
\frac{{{\beta}{\lambda_i}{(1-\rho)}D_0}}{R^{\textrm{UL}}(0,i)}}+\sum\limits^{p}_{i=1}{P_{\textrm{SR}}\frac{{{\beta}{\lambda_i}{(1-\rho)}D_0}}{R^{\textrm{UL}}(0,i)}}.
\end{split}$}
\end{eqnarray}

In summary, the total energy consumption of the swarm of drones can be represented as

\begin{eqnarray}
\resizebox{.75\hsize}{!}{$
\begin{split}
\label{zongnenghao}
E_{\textrm{Total}}&=E_{\textrm{Total}}^{\textrm{Comp}}+E^{\textrm{Trans}}_{\textrm{Total}}\\
&=kf_0^{\sigma}{\frac{\rho{\alpha_0}{D_0}}{f_0}}+ \sum\limits_{i=1}^{p}{kf_i^{\sigma}{ \frac{{{\alpha_0}{\lambda_i}{(1-\rho)}D_0}}{f_i}}} +\\
&\sum\limits^{p}_{i=1}{P_{\textrm{Tx}}
\frac{{{\beta}{\lambda_i}{(1-\rho)}D_0}}{R^{\textrm{UL}}(0,i)}}+\sum\limits^{p}_{i=1}{P_{\textrm{Rx}}\frac{{{\beta}{\lambda_i}{(1-\rho)}D_0}}{R^{\textrm{UL}}(0,i)}}.
\end{split}$}
\end{eqnarray}
\subsection{Problem Formulation}
To sum up, a problem to minimize the energy consumption of FCSD within latency and reliability constraints, is modeled as follows:
\begin{eqnarray}
\vspace{-0.15cm}
\label{p}
\mathcal{P:} \quad &(\rho,\lambda_i)=\arg \min E_{\textrm{Total}}
\end{eqnarray}
\begin{subnumcases}
{\textrm{s.t.}}
\label{con}
&\hspace{-2.5mm}$ \rho +\sum \limits_{i=1}^{p}\lambda_i(1-\rho)=1$\\
\label{con1}
&\hspace{-2.5mm}$T_{\textrm{Total}}  \leq  T_0 $ \\
\label{con2}
&\hspace{-2.5mm}$R_{\textrm{Total}}  \geq  R_0 $\\
\label{con3}
&\hspace{-2.5mm}$0\leq \lambda_i, \rho$
\end{subnumcases}
\vspace{-0.1cm}
\section{LRGA-MIE Algorithm}
To find the optimal solution of the problem $\mathcal{P}$, we design a latency and reliability constrained minimum energy consumption algorithm based on the real-code genetic algorithm (LRGA-MIE)\cite{optimization}.

Genetic algorithm (GA) is a kind of widely used heuristic algorithm due to its advantages of better global searching capability, strong robustness, parallel processing capability, etc. In the real-coded GA, each individual $\bm{X_{i}}=\{x_{i1},x_{i2},\cdots,x_{i(p+1)}\}$ in the population represents a possible solution of the optimization problem, which would be initially set to a random value. And then, through the constant evolution of selecting, crossing over and mutating the initial population, an optimal individual is found. Different from the unconstrained optimization problem, the problem $\mathcal{P}$ formulated has several constraints including equality and inequality constraints (i.e., Eq.~\eqref{con}, \eqref{con1}, \eqref{con2} and \eqref{con3}). However, GA cannot solve constrained optimization problem directly. Therefore, we adopt exterior penalty function method\cite{exterior} to transform the constrained problem into an unconstrained optimization problem.

In the following, the design details of LRGA-MIE algorithm are explained.

The fitness function of LRGA-MIE is reconstructed as follows:
\begin{equation}
\vspace{-0.15cm}
\resizebox{.89\hsize}{!}{$
\label{refitness}
f(\bm{X})=
\left\{
\begin{array}{l}
\hspace{-2pt}E_{\textrm{Total}}(\bm{X})\hspace{4.5cm} \bm{X}\in\bm{F};
\\[0.3cm]
\hspace{-2pt}E_{\textrm{Total}}(\bm{X})+h(g)\sum \limits_{j=1}^{p+4}E_{j}(\bm{X})+ \xi(\bm{X},g)\hspace{0.2cm} \bm{X}\in\bm{S-F},
\end{array}
\right.$}
\end{equation}
where $\bm{F}$ is the feasible region in the search space $\bm{S}$, and $\bm{S-F}$ denotes the infeasible region. $h(g)$ represents the penalty factor, which is a large number and usually taken a strictly increasing positive sequence that tends to infinity as the number of iterations increases. $E_{j}(\bm{X})$ is the constraint violation value of the infeasible individuals for the $jth$ constraint, and $\xi(\bm{X},g)$ indicates an additional heuristic value for infeasible individuals in the $gth$ generation. $E_{j}(\bm{X})$ and $\xi(\bm{X},g)$ can be expressed as
\begin{eqnarray}
\resizebox{.89\hsize}{!}{$
\label{yueshu}
E_{j}(\bm{X})=
\left\{
\begin{array}{l}
\max(0,-\bm{X}(j))\hspace{3.4cm} 1\leq j\leq p+1;
\\[0.3cm]
\mid\bm{X}{(1)} +\sum \limits_{i=2}^{p+1}\bm{X}{(i)}(1-\bm{X}{(1)})-1\mid \hspace{0.4cm}j=p+2;
\\[0.3cm]
\max(0,T_{\textrm{Total}}-T_{0}))\hspace{2.83cm}              j=p+3;
\\[0.3cm]
\max(0,R_{0}-T_{\textrm{Total}})) \hspace{2.77cm}            j=p+4,
\end{array}
\right.$}
\end{eqnarray}
\begin{equation}
\resizebox{.88\hsize}{!}{$
\label{qifashi}
\xi(\bm{X},g)= Wor(g)-\min\limits_{\bm{X}\in\bm{S-F}}\bigg\{E_{\textrm{Total}}(\bm{X})+h(g)\sum \limits_{j=1}^{p+4}E_{j}(\bm{X})\bigg\},$}
\end{equation}
respectively, where $E_{\textrm{Total}}(\bm{X})$ represents the fitness value of the $gth$ generation feasible individuals. $Wor(g)$ records the feasible individual with the worst fitness through $g$ generation evolution, and guarantee that the fitness of the feasible individuals are always better than the infeasible individuals during the course of the iteration. Whose value can be updated by
\begin{equation}
\resizebox{.78\hsize}{!}{$
\label{zaogaoshu}
Wor(g)=\max\left\{Wor(g-1),\max\limits_{\bm{X}\in\bm{F}}\{E_{\textrm{Total}}(\bm{X})\}\right\}.$}
\end{equation}

In the LRGA-MIE algorithm, each chromosome, namely each individual $X_{i}$ in the population is designed as a one-dimensional real array with $p+1$ genes, which should be randomly initialized with real number in the searching space $\bm{S}$ firstly. Then, the fitness value of each individual would be calculated according to Eq.~\eqref{refitness} to evaluate the population. Next, the genetic operators are performed to update the initial population. And the specific genetic operators are given as follows:

Selection: In this paper, 2-tournament selection strategy with elitism preservation\footnote{Analyzing the convergence and the time complexity of GA is an extremely challenging theoretical issue in the evolutionary computation area, which is beyond the scope of this paper. But it has been proved that the GA with elitism preservation must converge to the global optimal solution \cite{eiben1990global}.} is adopted for its advantages of simplicity and efficiency. Firstly, the individuals with the lowest fitness values (i.e., elitism individuals) are directly retained into the next generation of populations. Then, the remaining individuals are randomly selected in pairs and the individual with lower fitness value will be retained to the next generation.

Crossover: Crossover is to passed the original good genes onto the offspring. Where two new children individuals (i.e., $\bm{X'_{1}}$, $\bm{X'_{2}}$) are generated  by a linear combination of the two parent individuals (i.e.,~$\bm{X_{1}}$, $\bm{X_{2}}$). The relationship between offspring and parents can be described as
\vspace{-0.5mm}
\begin{equation}
\resizebox{.45\hsize}{!}{$
\left\{
             \begin{array}{lr}
                \label{EQ:22}
                \hspace{-1.5mm}\bm{X'_{1}}=\delta\bm{X_{1}}+(1-\delta)\bm{X_{2}};\\
               \\[-0.4mm]
                \hspace{-1.5mm}\bm{X'_{2}}=\delta\bm{X_{2}}+(1-\delta)\bm{X_{1}},
             \end{array}
\right.
$}
\vspace{-0.5mm}
\end{equation}
respectively, where $\delta$ is a random number on interval $(0,1)$.

Mutation: Mutation operation determines the local search capability of the LRGA-MIE and improves the diversity of individuals in the population. In this paper, the non-uniform mutation operator is applied.
When the individual $\bm{X_{i}}=\{x_{i1},x_{i2},\cdots,x_{il},\cdots,x_{i(p+1)}\}$ mutates into the new individual $\bm{X'_{i}}=\{x_{i1},x_{i2},\cdots,x'_{il},\cdots,x_{i(p+1)}\}$, the new gene $x'_{il}$ can be calculated as
\begin{equation}
\resizebox{.88\hsize}{!}{$
\label{EQ:24}
x'_{il}\!\! =\!\!
\left\{
\begin{array}{l}
\!\!\!\!x_{il}+(1-x_{il})(1-q^{(1-g/G)b})\hspace{0.2cm} \textrm{if}\hspace{0.2cm} random(0,1)=0,
\\[0.3cm]
\!\!\!\!x_{il}+x_{il}(1-q^{(1-g/G)b})\hspace{1.05cm}  \textrm{if}\hspace{0.2cm} random(0,1)=1.
\end{array}
\right.$}
\end{equation}
Where $q$ is a random number in the range of $[0,1]$ with uniform distribution. $g$ is the current evolution generation, and $G$ represents the maximum evolution generation. $b$ is a system parameter, which determines the degree of dependence of the random number perturbations on the evolution generation $g$. The value of $b$ is in the range of $[2,5]$. $random(0,1)$ denotes any value of 0 or 1 with equal probability.

To minimize the energy consumption in Eq.~\eqref{zongnenghao}, the basic steps of LRGA-MIE are shown in Algorithm \ref{alg:RCGA}.

The time complexity of LRGA-MIE can be presented by $\mathcal{O}\left(G*S*(p+1)\right)$, where $S$ represents the population size. According to \cite{zhu2017discrete}, $S$ and $G$ are linear functions with respect to $p+1$, hence the time complexity of LRGA-MIE is $\mathcal{O}\left(p^{3}\right)$.

\begin{algorithm}[h]
\setlength{\abovecaptionskip}{-0.1cm}   
  \setlength{\belowcaptionskip}{-0.1cm}   
\caption{LRGA-MIE algorithm}
\label{alg:RCGA}
\begin{algorithmic}[1]\scriptsize

\REQUIRE
$p$, $\mu_{0,i}$, $\nu_{i}$, $\nu_{0}$, $\mathcal{D}$, $\mathcal{F}$, ${\varPsi_0}$, $G$, $S$\\     ~~~~~$pc$: Crossover probability, $pm$: Mutation probability.
\ENSURE
BestFitness, BestSolution
\STATE Randomly initialize each individual $\bm{X_{i}}$ in the Population
\STATE globalBestFitness = 0
\FOR{Generation:1 to $G$}
  \STATE localBestFitness = 0
\FOR{each individual $\bm{X_{i}}$ $\in$ Population}
    \STATE Calculate the value of $E_{j}(\bm{X})$ using equation \eqref{yueshu}
\ENDFOR
  \STATE Calculate the the value of $\xi(\bm{X},g)$ and $Wor(g)$ using equation \eqref{qifashi} and \eqref{zaogaoshu}
\FOR{each individual $\bm{X_{i}}$ $\in$ Population}
  \STATE Calculate the fitness value $f(\bm{X_{i}})$ using equation \eqref{refitness}
\IF{$f(\bm{X_{i}})$ $<$ localBestFitness}
  \STATE localBestFitness = $f(\bm{X_{i}})$
  \STATE localBestSolution = $\bm{X_{i}}$
\ENDIF
\ENDFOR
\IF{localBestFitness $<$ globalBestFitness}
\STATE globalBestFitness = localBestFitness
\STATE BestSolution = localBestSolution
\ENDIF
\STATE Select individuals from the Population;
\IF{rand $<$ $pc$}
\STATE crosspop = cossover(Population, $pc$)
\ENDIF
\IF{rand $<$ $pm$}
\STATE mutatepop = mutate(crosspop, $pm$)
\ENDIF
\STATE Update the Population: Population = mutatepop
\ENDFOR
\label{code:RCGA:Calculate4}\\
\RETURN BestSolution, globalBestFitnes
\end{algorithmic}
\end{algorithm}
  \vspace{-3mm}
\section{performance and evaluation}
In this section, to testify the performance of the FSCD system with LRGA-MIE algorithm, a set of simulation results are presented. Referring to\cite{chuanshusulv,dinh2017offloading,qi2017optimal,cheng2016optimal,jiang2013renewal,jiang2013joint,zhu2017non}, the system parameters of FCSD are summarized in TABLE \ref{default_para}.

\begin{table}
\setlength{\abovecaptionskip}{-0.1cm}   
  \setlength{\belowcaptionskip}{-0.1cm}   
  \centering
  \caption{System parameters of FCSD}\label{default_para}
  \scalebox{0.70}{
  \begin{tabular}{|c|c||c|c|}
  \hline
  \textbf{Parameter} & \textbf{Value} & \textbf{Parameter} & \textbf{Value}  \\
  \hline
  $W^{\textrm{UL}}$ & 1 MHz & $f_0, f_i $ & $\textrm{Unif([0.2, 0.9] GHz)}$ \\
  \hline
  $N_0$ & -100 dBm & $(x_{0},y_{0},z_{0})$ & $\textrm{(0 m, 0 m, 0 m)}$\\
  \hline
  $P_{\textrm{Tx}}$ & 1.258 W & $(x_{i},y_{i},z_{i})$ & randomly in 100 $\textrm{m}^{3}$ area \\
    \hline
  $P_{\textrm{Rx}}$ & 1.181 W &  $ \nu_{0}, \nu_{i}$ & $\textrm{Unif([0.001, 0.3])}$ \\
  \hline
  $\gamma$ & 3 &  $\mu_{i}$ & $\textrm{Unif([0.001, 0.3])}$ \\
  \hline
  $h_0$ & CN(0,1) &   $f_{c}$ & 1 GHz \\
  \hline
  $k$ & $1.25 \times10^{-26}$ & $W^{c}$ &  2 MHz\\
  \hline
  $\sigma$ & 3 & $\mu _{c}$ & 0.17 \\
  \hline
  $r$ & $100 \hspace{0.1cm} m^{3}$ &  $\nu_{c}$  & 0.00001 \\
  \hline
  $\beta$ & 1 &   $(x_{c}, y_{c}, z_{c})$ & $\textrm{(2000 m, 2000 m, 2000 m)}$ \\
  \hline
  \end{tabular}}
  \vspace{-5mm}
\end{table}
The parameters of the algorithm are set as follows: The maximum number of iterations is 300. The population size (i.e., $G$) is 100. The crossover and mutation probability (i.e., $pc$ and $pm$) are set as 0.8 and 0.1, respectively. And the value of $Wor(0)$ is set as $10^{5}$.

In the following simulations, the parameters of the task ${\varPsi_0}$ are set as follows, unless otherwise specified or used as variables. $D_0$ is set as 1 MB. $\alpha_{0}$ is set as 1900/8 to represent a computational intensive task\cite{power2}. $T_{0}$ and $R_{0}$ are set as 0.8 s and 0.99, respectively. And we assumed that there are a total of 10 drones available nearby that can serve as fog computing nodes to help $dr_0$ achieved the computing task $\varPsi_0$, i.e., $p=10$. The numerical results in this section are based on an average value over 3000 Monte Carlo simulations.
\vspace{-3mm}
\subsection{Latency Performance Comparison of Three Computation Architectures }
\begin{figure}[htb]
 \vspace{-0.2cm}
\setlength{\abovecaptionskip}{-0.1cm}   
  \setlength{\belowcaptionskip}{-1cm}   
  \centering
  \includegraphics[width=5.3cm]{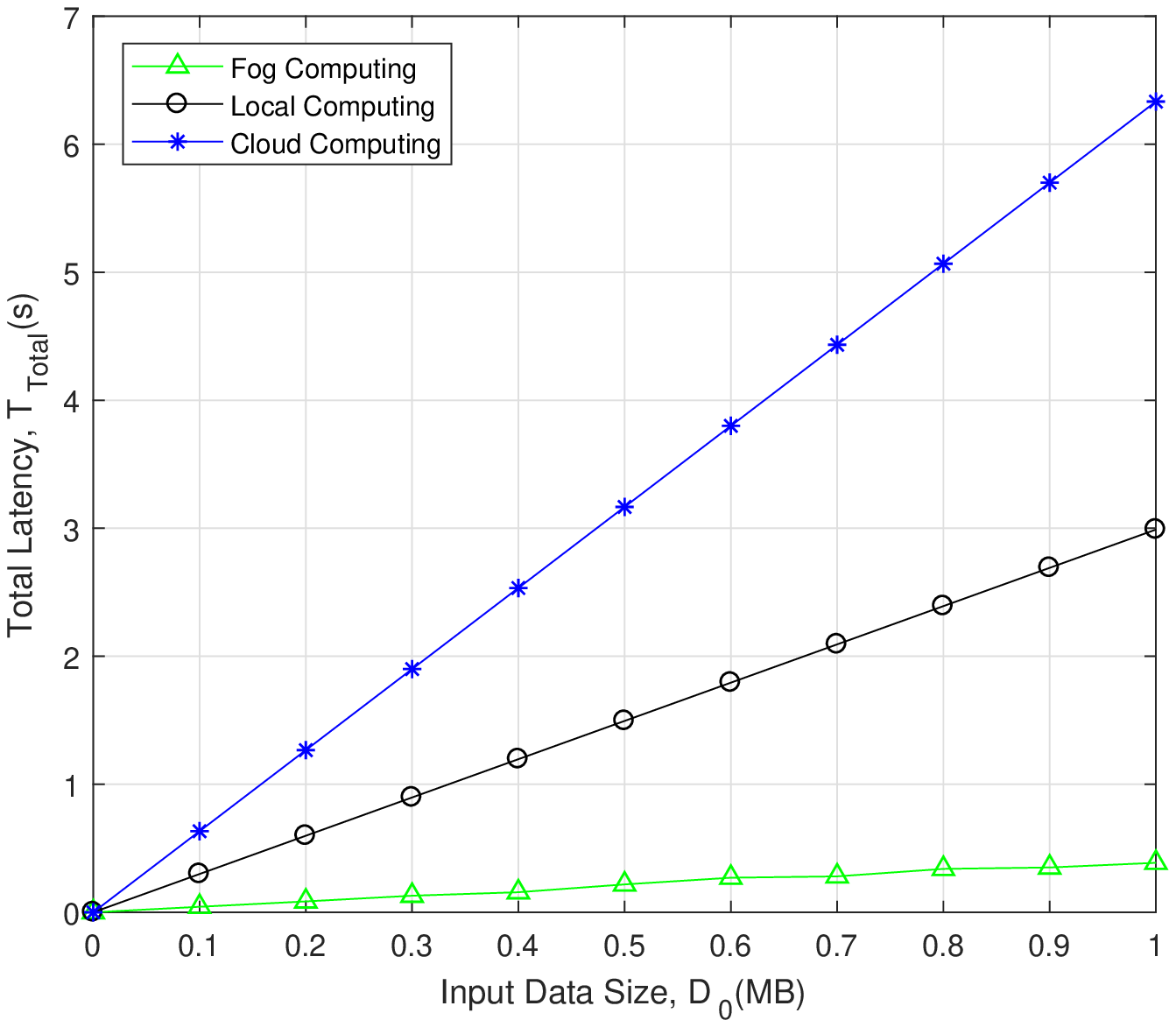}
  \caption{Latency performance comparison of three computation architecture}\label{latency1}
     \vspace{-0.2cm}
\end{figure}

Fig. \ref{latency1} shows the latency performance comparison of three computation architectures, i.e., cloud computing, local computing and fog computing. With the increasing of the input data size, the cloud computing curve is intensely higher than the local computing curve and fog computing curve. The reason is that the transmission latency of cloud computing architecture is growing linearly with the increasing of the input data size, due to the long distance data transmission and limited bandwidth between cloud and the drone $dr_0$. Furthermore, the local computing curve is relatively lower than the cloud computing curve but relatively higher than the fog computing curve. This is because that the drone $dr_0$ has a certain amount of computing ability. When the computing task is relatively smaller, it is able to handle the task locally in a relatively low computation latency and without transmission latency. But, when the input data size increased, limited by its computing ability, the local computing manner cannot complete the computing task with lower latency. The fog computing latency, as we can see, is always lower than other two kinds of computing manners, this is because that the transmission latency of fog computing is relatively lower due to the surrounding fog nodes are intensely close, and meantime the computation latency is relatively lower as well, because of it integrates the computational ability of numerous fog nodes. When the input data size of the task $\varPsi_0$ is 0.5 MB, we can observe that the latency performance of fog computing improved by 93.12\% and 85.42\% compared with the cloud computing and local computing respectively. Therefore, the fog computing based computation offloading is suitable for latency sensitive business of swarm of drones.
\vspace{-0.3cm}
\subsection{Reliability Performance Comparison of Different Algorithms}
\begin{figure}[htb]
\vspace{-5mm}
\setlength{\abovecaptionskip}{-0.1cm}   
 \setlength{\belowcaptionskip}{-0.1cm}   
\centering
  \includegraphics[width=5.3cm]{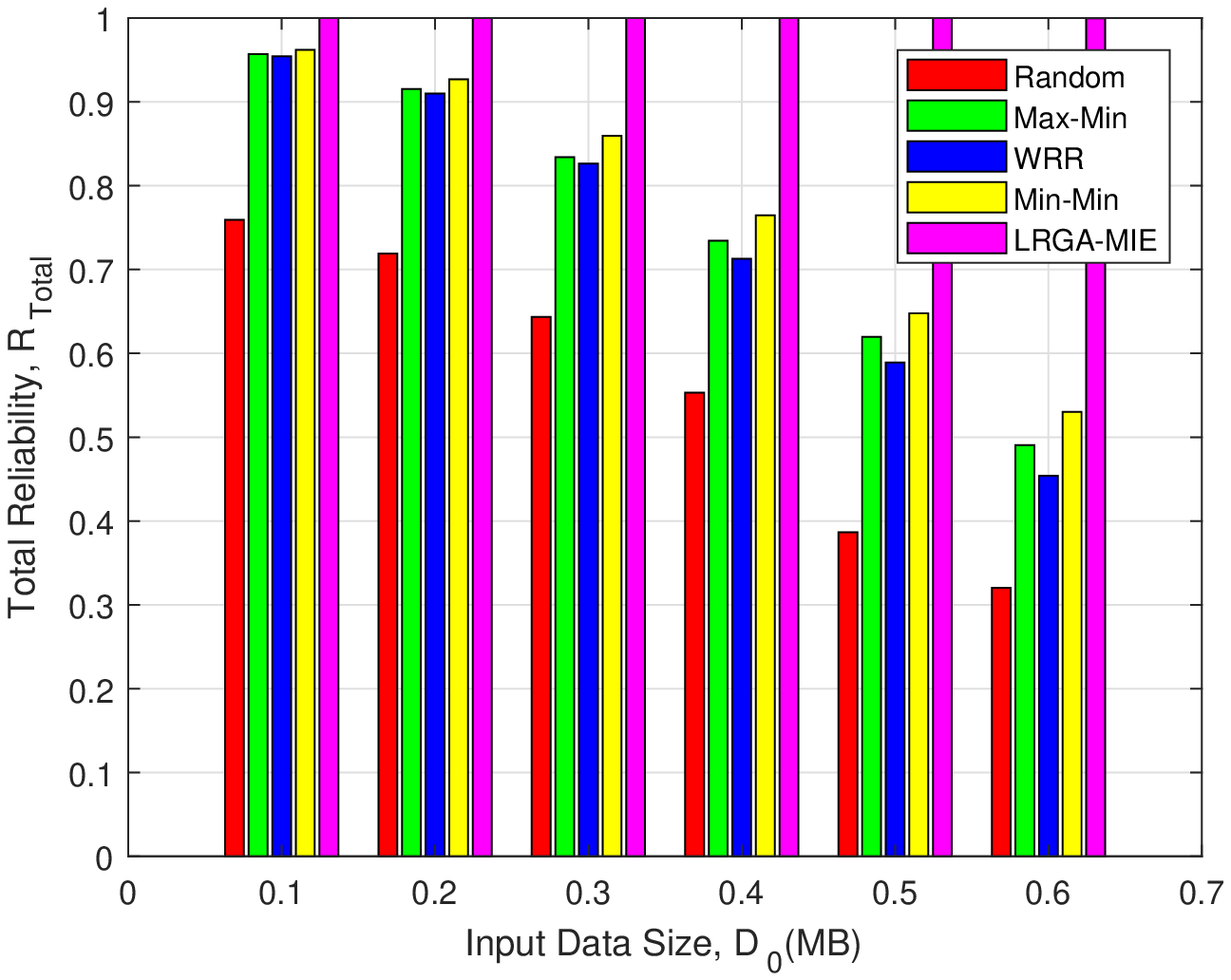}
  \caption{Reliability performance comparison of different algorithms}\label{Reliability}
 \vspace{-0.2cm}
\end{figure}

In this section, we analyze the high efficiency of the LRGA-MIE algorithm in improving reliability in FCSD by comparing it with the Max-Min\cite{maxmin}, Weighted Round Robin (WRR)\cite{wrr} and Min-Min\cite{minmin} algorithm, and also the random task assignment. The simulation results are shown in Fig. \ref{Reliability}. As we can see, when the input data size is relatively smaller, the Max-Min, WRR, Min-Min and LRGA-MIE algorithms all have good performance, compared to the random task assignment. Specifically, when the input data size is 0.1 MB, the reliability of these algorithms are all higher than 0.95. However, with the increasing of the input data size, the reliability performance of the Max-Min, WRR, Min-Min, LRGA-MIE algorithms decrease in different degree. This is because the increase of input data size will increase the processing latency of some fog nodes and even the entire FCSD system, according to Eq.~\eqref{eq8}. Furthermore, no matter which fog nodes increase in transmission latency or computation latency, the total reliability of the FCSD system will be decreased, according to Eq.~\eqref{zongnenghao}. But for LRGA-MIE algorithm, as we can see from the Fig.~\ref{Reliability}, it has always maintains intensely higher reliability. As the input data size increase from 0.2 MB to 0.6 MB, the reliability of LRGA-MIE only decrease from 0.9997 to 0.9989. And when the input data size is 0.6 MB, we can observe that the reliability of LRGA-MIE is higher than that of Max-Min, WRR, Min-Min algorithm and random task allocation strategy by 0.4674, 0.529, 0.5041 and 0.683, respectively. Therefore, the LRGA-MIE algorithm is better suited for optimizing reliability with global consideration of computation capability, transmission capability and failure rate.

\vspace{-0.3cm}
\subsection{Energy Consumption Performance Impact by Latency Constraint and Reliability Constraint}
\begin{figure}[htb] 
\setlength{\abovecaptionskip}{-0.1cm}   
 \setlength{\belowcaptionskip}{-0.3cm}   
\centering
  \includegraphics[width=5.3cm]{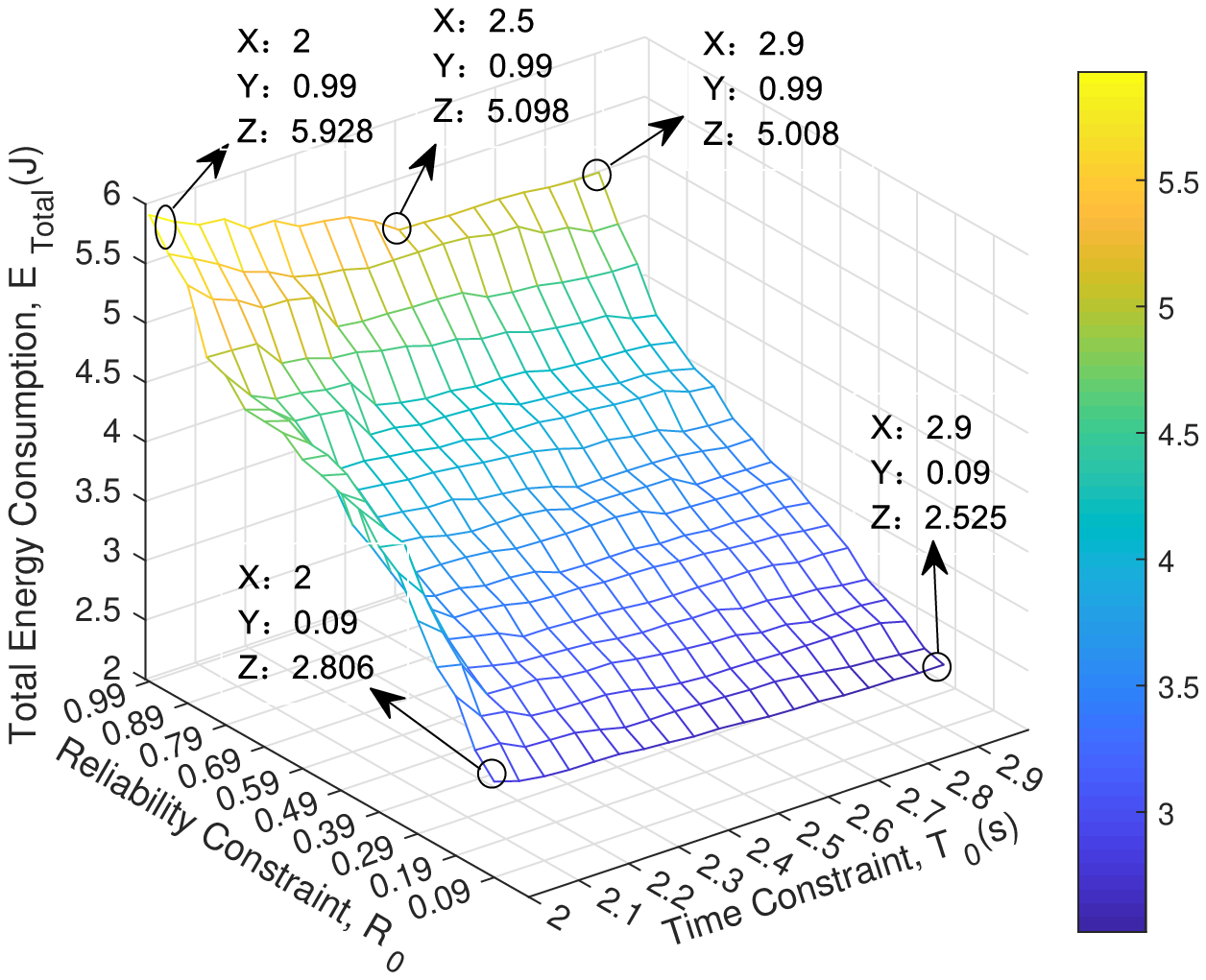}
  \caption{Energy consumption performance impact by latency constraint and reliability constraint}\label{3d}
   \vspace{-0.2cm}
\end{figure}

In this section, we discuss the impact of latency and reliability constraints on total energy consumption of FCSD. The simulation results are shown in Fig. \ref{3d}. As we can see, when the latency constraint is fixed, the total energy consumption will gradually decrease with the reducing of reliability constraint. Similarly, when the reliability constraint is fixed, the total energy consumption will gradually decrease with the improving of latency constraint. This is because whether the reliability constraint is reduced or the latency constraint is increased, the feasible domain of the optimization problem is extended, and thus more solutions with low energy consumption performance can be obtained. However, we can observe that when the reliability is fixed, e.g., $R_0=0.99$, as the latency constraint increases gradually, the energy consumption curve will drop rapidly first, i.e., $T_0$ ranges from 2 s to 2.5 s, then the rate of decline will slow down and finally the curve will be tend to be gentle, i.e., $T_0$ ranges from 2.5 s to 2.9 s. It is because that, in this reliability constraint, the computing task $\varPsi_0$ can be completed within 2.5 s. When the latency constraint is greater than 2.5 s, the latency constraint is not the main factor hindering the performance of the system, therefore, the increasing of the latency constraint will not have a significant impact on the energy consumption performance of FCSD.
\vspace{-0.3cm}
\subsection{Energy Consumption Performance Comparison of Different Algorithms}
\begin{figure}[htb!]
\setlength{\abovecaptionskip}{-0.1cm}   
\centering
  \includegraphics[width=5.3cm]{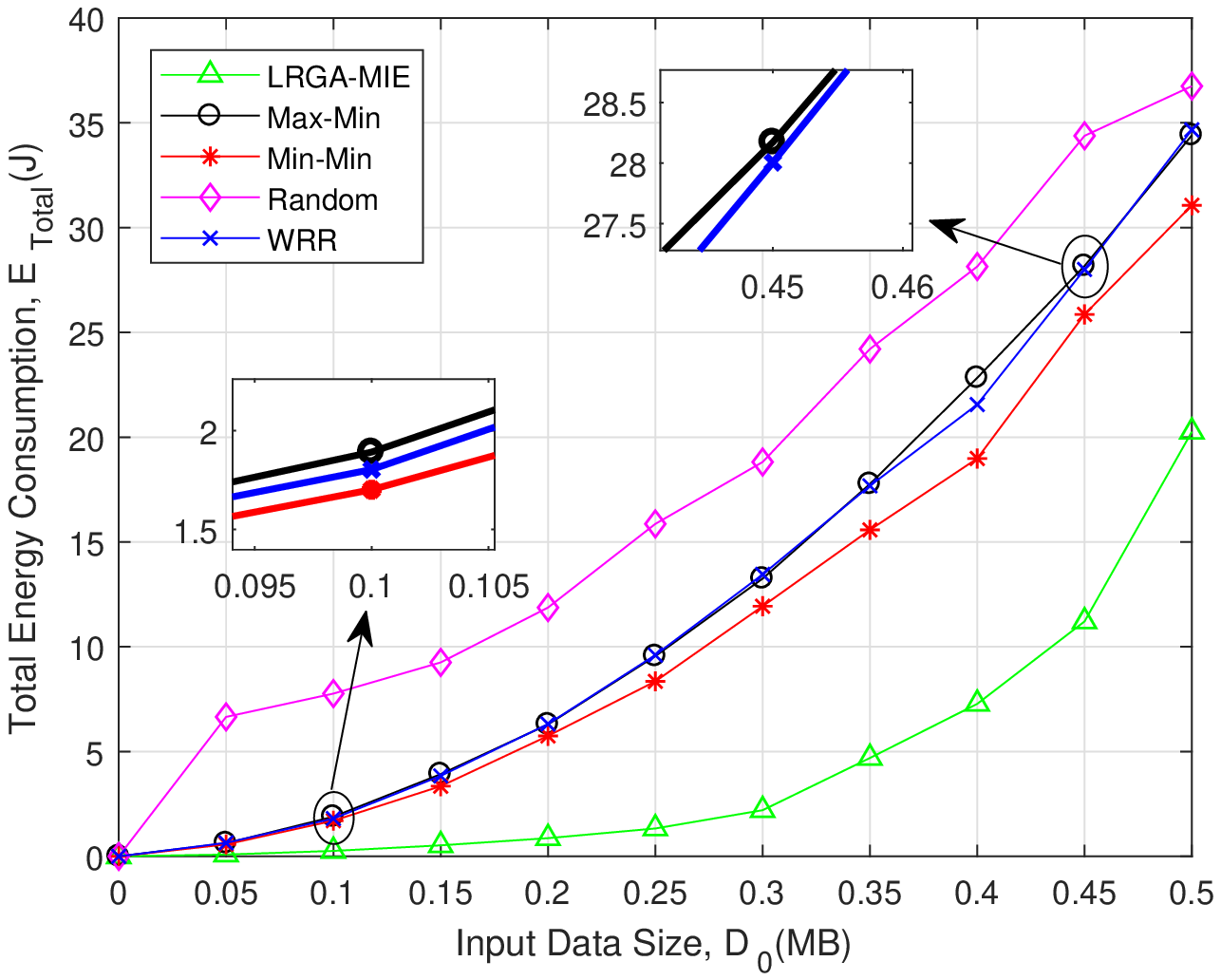}
  \caption{Energy consumption performance of different algorithms}\label{Energy_al_D}

\end{figure}
In this section, we analyze the energy consumption performance of different algorithms in FCSD. The simulation results are shown in Fig. \ref{Energy_al_D}. We can observe that the random task assignment has the worst energy consumption performance. When the input data size is relative smaller, although not as good as LRGA-MIE algorithm, the performance of the WRR, Max-Min and Min-Min algorithms are not bad, compared to the random task assignment. However, with the increasing of the input data size, the energy consumption of the Max-Min, Min-Min and WRR algorithms and random task assignment increase rapidly. As we can see, the energy consumption of Max-Min algorithm approaches that of the WRR algorithm, and the Min-Min algorithm is slighter better than the Max-Min and WRR algorithms. But the LRGA-MIE algorithm maintains good performance all along, due to its strong global search ability. It shows that under the premise of ensuring low latency and high reliability requirements of the computing task, the LRGA-MIE algorithm has intensely good performance in reducing the energy consumption in FCSD, and it has strong adaptability and stability to the growth of the input data size. When the input data size is 0.5 MB, the energy consumption performance of the LRGA-MIE algorithm improved by 44.87\%, 41.16\%, 41.12\% and 34.77\%, compared with the random task assignment, Max-Min, WRR and Min-Min algorithm, respectively.
\vspace{-1mm}
\section{Conclusion}
In this paper, to solve the problem that the cloud based computation offloading is not suitable for addressing the latency and reliability sensitive task, we introduced the fog computing based computation offloading into swarm of drones (FCSD). And specifically focusing on the latency and reliability business scenarios and improving the practicality of FCSD, an optimization problem to minimize the energy consumption of FCSD within latency and reliability constraints is constructed. In order to solve the NP-hard problem we formulated, the LRGA-MIE algorithm was proposed. The simulation results demonstrated that the LRGA-MIE can minimize the energy consumption of FCSD, on the basis of completing the computing task within latency and reliability requirements. In future research work, two things are on our agenda. One is to reduce the algorithm complexity of LRGA-MIE to further improve its practicability. The other is to utilize some new technologies (e.g., cognitive radio\cite{haykin2005cognitive}, orbital angular momentum\cite{cheng2018orbital}, etc.) to solve the problem of spectrum resource shortage when the number of UAV increases rapidly.
\vspace{-1mm}
%
%

\bibliography{ref}

%
%

\end{document}